\documentclass[journal]{IEEEtran}
\ifCLASSINFOpdf
\else
\fi

\usepackage{amssymb}
\usepackage{latexsym}
\usepackage{graphicx}
\usepackage{color}
\usepackage{verbatim}
\usepackage{multirow}
\usepackage{amsmath}
\usepackage{amsfonts}

\newcommand{\E}{\ensuremath{\mathbb E}}

\newcommand{\R}{\ensuremath{\mathbb R}}

\newcommand{\abu}{{\bf a}}

\newcommand{\kbu}{{\bf k}}

\newcommand{\nbu}{{\bf n}}
\newcommand{\sbu}{{\bf s}}

\newcommand{\rbu}{{\bf r}}

\newcommand{\xbu}{{\bf x}}
\newcommand{\ybu}{{\bf y}}

\newcommand{\Cbu}{{\bf C}}
\newcommand{\Dbu}{{\bf D}}

\newcommand{\Ibu}{{\bf I}}

\newcommand{\Sbu}{{\bf S}}

\newcommand{\Ubu}{{\bf U}}

\newcommand{\Phibu}{{\bf \Phi}}

\newcommand{\Psibu}{{\bf \Psi}}

\newcommand{\btheta}{{\boldsymbol \theta}}

\newcommand{\iproof}{{\noindent \textit{Proof}}}

\newcommand{\qed}{\hfill \ensuremath{\Box}}

\newtheorem{prop}{Proposition}
\newtheorem{df}{Definition}
\newtheorem{thr}{Theorem}
\newtheorem{lem}{Lemma}

\newtheorem{cor}{Corollary}

\numberwithin{const2}{const}

\begin{document}
%
\title{Indistinguishability and Energy Sensitivity of
Asymptotically Gaussian Compressed Encryption}
%
%
%

\author{Nam Yul Yu,~\IEEEmembership{Member,~IEEE}
\thanks{The author is with the School of
of Electrical Engineering and Computer Science (EECS), Gwangju Institute of Science and Technology (GIST), Korea.
(e-mail: nyyu@gist.ac.kr).}  
}

\maketitle

\begin{abstract}
The principle of compressed sensing (CS) can be applied in a cryptosystem
by providing the notion of security.
In information-theoretic sense, it is known that a CS-based cryptosystem can be perfectly secure
if it employs a random Gaussian sensing matrix updated at each encryption
and its plaintext has constant energy.
In this paper,
we propose a new CS-based cryptosystem that employs a secret bipolar keystream
and a public unitary matrix,
which can be suitable for practical implementation
by generating and renewing the keystream in a fast and efficient manner.
We demonstrate that the sensing matrix is asymptotically Gaussian for a sufficiently large
plaintext length,
which guarantees a reliable CS decryption for a legitimate recipient.
By means of probability metrics, we also show
that the new CS-based cryptosystem can have the indistinguishability against an adversary,
as long as the keystream is updated at each encryption and each plaintext has constant energy.
Finally, we investigate how much the security of the new CS-based cryptosystem
is sensitive to energy variation of plaintexts.
\end{abstract}

\begin{IEEEkeywords}
Compressed encryption,
Hellinger distance,
indistinguishability,
linear feedback shift register (LFSR),
probability metrics,
self-shrinking generators,
total variation distance.
\end{IEEEkeywords}

%
\IEEEpeerreviewmaketitle

\section{Introduction}

Compressed sensing (CS)~\cite{Donoho:CS}$-$\cite{CanTao:univ} is to recover a sparse signal
from the measurements that are believed to be incomplete.
A signal $\xbu \in \R^N$ is called $K$-\emph{sparse} 
if it has at most $K$-nonzero entries,
where $K \ll N$.
A sparse signal is linearly measured by $\ybu = \Phibu \xbu  \in \R^M $,
where $\Phibu$ is an $M \times N$ sensing matrix with $M \ll N$.
In CS theory, if 
$\Phibu$ obeys the \emph{restricted isometry property (RIP)}~\cite{Donoho:CS}\cite{CanTao:univ},
a stable and robust reconstruction of $\xbu$ can be guaranteed
from the incomplete measurement $\ybu$.
The CS reconstruction 
is accomplished by solving an $l_1$-minimization problem
with convex optimization
or greedy algorithms~\cite{Eldar:CS}.
With efficient measurement and stable reconstruction,
the CS technique has been of interest in
a variety of research fields, e.g., communications~\cite{Tropp:beyond}$-$\cite{Haupt:toep}, sensor networks~\cite{Duarte:dist}$-$\cite{Caione:wsn},
image processing~\cite{Duarte:single}$-$\cite{Lustig:mr}, radar~\cite{Gog:radar}, etc.

The CS principle can be applied in a cryptosystem for information security.
A \emph{CS-based cryptosystem} encrypts
a plaintext through a CS measurement process,
where the sensing matrix is kept secret.
The ciphertext can then be decrypted through a CS reconstruction process
by a legitimate recipient with the knowledge of the sensing matrix.
In~\cite{Baron:sec}, Rachlin and Baron proved that
a CS-based cryptosystem cannot be perfectly secure in itself, but might be computationally secure.
Orsdemir \emph{et al.}~\cite{Ors:secrub} showed that it is computationally secure against
a key search technique via an algebraic approach.
In~\cite{Bianch:linear}, Bianchi \emph{et al.} analyzed
the security of a CS-based cryptosystem
employing a random Gaussian sensing matrix updated at each encryption.
Precisely, they showed that the cryptosystem with the \emph{one-time sensing} random Gaussian matrix
can be perfectly secure, as long as each plaintext has constant energy.
A similar analysis has been made for a CS-based cryptosystem
having a circulant sensing matrix for efficient CS processes~\cite{Bianch:circ}\cite{Bianch:anal}. 
In~\cite{Reeves:wiretap} and \cite{Agrawal:wiretap}, 
wireless channel characteristics could be exploited
for wireless security of CS-based cryptosystems.
The CS technique can also be applied
in database systems~\cite{Li:privacy}, where
random noise has been intentionally added to CS measurements for differential privacy.
In practice, a variety of CS-based cryptosystems concerning the security
of multimedia, imaging, and smart grid data
have been suggested in~\cite{Dau:establish}$-$\cite{Zhang:review}.

%

In this paper, we propose a new CS-based cryptosystem that employs a secret bipolar keystream
and a public unitary matrix,
which can be suitable for practical implementation
by generating and renewing the keystream at each encryption
in a fast and efficient manner.
The keystream generator, based on a linear feedback shift register (LFSR),
plays a crucial role in the efficient implementation.
We demonstrate that the entries of the sensing matrix are asymptotically Gaussian distributed
if the plaintext length is sufficiently large.
With the sensing matrix, it is obvious that the new CS-based cryptosystem,
named as the \emph{asymptotically Gaussian one-time sensing (AG-OTS)} cryptosystem,
theoretically guarantees
a stable and robust CS decryption for a legitimate recipient.

For security analysis,
we study the indistinguishability~\cite{Katz:modern} of the AG-OTS cryptosystem. 
The \emph{total variation (TV)} distance~\cite{Gibbs:bounding}\cite{LeCam:asymp}
between probability distributions of ciphertexts conditioned on a pair of plaintexts
is examined as a security measure
for the indistinguishability, where
the upper and lower bounds on the TV distance
are developed by the \emph{Hellinger} distance~\cite{Gibbs:bounding}\cite{LeCam:asymp}.
With the probability metrics,
we examine the success probability of an adversary to distinguish a pair of potential plaintexts from
a given ciphertext.
By proving that the success probability of any kind of attack is at most that of a random guess,
we demonstrate that
the AG-OTS cryptosystem can have the indistinguishability, 
as long as 
each plaintext has constant energy.
Therefore,
the AG-OTS cryptosystem, if it has a normalization step before encryption
for equalizing the plaintext energy, can be computationally secure.

Finally, we investigate how much the security of the AG-OTS cryptosystem
is sensitive to energy variation of plaintexts.
It is worth studying the energy sensitivity,
since one might need to assign unequal energy for plaintexts in the presence of noise,
depending on the reliability demands.
As a consequence, we develop sufficient conditions on the minimum energy ratio, the plaintext length, and the
maximum plaintext-to-noise power ratio,
respectively,
to achieve the \emph{asymptotic} indistinguishability of the AG-OTS cryptosystem
having unequal plaintext energy.
Since the analysis relies on the Gaussianity of the sensing matrix,
the results of the energy sensitivity
can also be applicable to the Gaussian one-time sensing (G-OTS) cryptosystem in~\cite{Bianch:linear}.




This paper is organized as follows.
In Section II, we propose a new CS-based cryptosystem
employing a secret bipolar keystream,
where the sensing matrix turns out to be asymptotically Gaussian.
Also, we discuss an LFSR-based efficient keystream generation for the cryptosystem.
Section III introduces the indistinguishability along with
the probability metrics of total variation (TV) and Hellinger distances.
For security analysis,
Section IV studies the indistinguishability and the energy sensitivity of the new CS-based cryptosystem
in the presence of noise.
Section V presents numerical results to demonstrate the security of the new CS-based cryptosystem.
Finally, concluding remarks will be given in Section VI.

%
%
%
%

\emph{Notations}: 
A matrix (or a vector) is represented by
a bold-face upper (or lower) case letter.
$\Ubu^T$ and $|\Ubu|$ denote the transpose and the determinant
of a matrix $\Ubu$, respectively.
$\Ubu(k, t)$ is an entry of an $M \times N$ matrix $\Ubu$ in the $k$th row and the $t$th column,
where $0 \leq k \leq M-1$ and $0 \leq t \leq N-1$.
Also, $\Ubu(k, :)$ denotes the $k$th row vector of $\Ubu$,
while $\Ubu(:, t)$ is the $t$th column vector of $\Ubu$.
${\rm diag} (\sbu)$ is a diagonal matrix whose diagonal entries are from a vector $\sbu$.
An identity matrix is denoted by $\Ibu$,
where the dimension is determined in the context.
$\Dbu$ denotes an $N \times N$ discrete-cosine transform (DCT) matrix,
where $\Dbu \Dbu^T = \Dbu^T \Dbu = N \Ibu$.
For an $N$-dimensional vector $\xbu = (x_1,\cdots, x_{N})^T \in \R^N$, the $l_p$-norm of $\xbu$ is denoted by
$ || \xbu ||_{p} = \left( \sum_{k=1} ^{N} |x_k|^p \right) ^{\frac{1}{p}} $, where
$1 \leq p < \infty $. If the context is clear, $|| \xbu ||$ denotes the $l_2$-norm of $\xbu$.
A vector $\nbu \sim {\mathcal  {N}} ({\bf 0}, \sigma^2 \Ibu)$ is a
Gaussian random vector
with mean ${\bf 0} = (0, \cdots, 0)^T$ and covariance $\sigma^2 \Ibu$.
Finally, $\E[\cdot]$ denotes the average of a random vector or a random matrix.

\section{System Model}

In~\cite{Bianch:linear}, the authors presented the Gaussian one-time sensing (G-OTS) cryptosystem,
where a random Gaussian sensing matrix is used only once for each encryption, and renewed for the next.
In information-theoretic sense, 
they showed that if each plaintext has constant energy,
the G-OTS cryptosystem can be perfectly secure, 
which implies the \emph{indistinguishability}~\cite{Katz:modern}
that will be discussed in next section.

In practice, generating the Gaussian entries at each encryption
may require high complexity and large memory for CS encryption and decryption.
For efficient implementation,
this section proposes a new CS-based cryptosystem in which
the sensing matrix employs a bipolar keystream.

\subsection{Asymptotically Gaussian Sensing Matrices}

\begin{df}\label{def:Phinew}
Let $\Ubu \in \R^{N \times N} $ be a public unitary matrix,
i.e., $\Ubu^T \Ubu = \Ubu \Ubu ^T = N \Ibu$,
where each element of $\Ubu$ has the magnitude of ${\mathcal O} (1)$.
Let $\Sbu $ be a secret $M \times N$ matrix, where we assume that
each element takes $\pm 1$ independently and uniformly at random.
Then, a new CS-based cryptosystem has the sensing matrix of
\begin{equation*}\label{eq:circ}
\Phibu = \frac{1}{\sqrt{MN}}  \Sbu \Ubu.
\end{equation*}
\end{df}

Theoretically, each element of $\Sbu$ can be taken from the random Bernoulli distribution.
In practice, however, we consider a keystream generator of stream ciphers
to generate it in a fast and efficient manner.
Employing an efficient keystream generator allows us to
construct and update $\Sbu$ at each encryption with low complexity and small memory.
Since a keystream for a stream cipher is designed to have nice pseudorandomness properties~\cite{GolGong:SD},
such as balance, large period, low autocorrelation, large linear complexity, etc.,
we assume that each entry of $\Sbu$ from the keystream
takes $\pm 1$ independently and uniformly at random, which
facilitates the reliability and security analysis of the new CS-based cryptosystem.


\begin{thr}\label{th:aGauss}
In Definition~\ref{def:Phinew},
the elements of $\Phibu$ follow the Gaussian distribution asymptotically
for a sufficiently large $N$.
\end{thr}

\iproof:
Each row of $\Phibu$ is represented by
\begin{equation}\label{eq:phibu_a}
\begin{split}
\Phibu(k, :) & = \frac{1}{\sqrt{MN}} \Sbu(k, :) \Ubu \\
& = \sqrt{\frac{N}{M}} \cdot \frac{\bf 1}{\sqrt{N}} {\rm diag}(\Sbu(k, :)) \frac{\Ubu}{\sqrt{N}} ,
\quad k =1 ,\cdots, M
\end{split}
\end{equation}
where $\Sbu(k, :) $ is the $k$th row vector of $\Sbu$, and
${\bf 1} = (1, \cdots, 1)$ is all one row vector of length $N$, respectively.
In~\eqref{eq:phibu_a}, $\frac{\bf 1}{\sqrt{N}}$
is a row of a unit-norm row matrix
with absolute magnitude of all entries of ${\mathcal O}\left(\frac{1}{\sqrt{N}} \right)$.
Also, $\frac{\Ubu}{\sqrt{N}}$ is a unit-norm column matrix with the maximum
absolute magnitude of entries of $o(1)$.
With the structure,
Theorem III.1 of \cite{Do:fast} shows that
the elements of $\Phibu(k, :)$ are asymptotically Gaussian if $N$ is sufficiently large,
which completes the proof.
\qed

The asymptotic Gaussianity of Theorem~\ref{th:aGauss} also holds
if the elements of $\Sbu$ are generated by an efficient keystream generator,
under the assumption that each one 
takes $\pm 1$ independently and uniformly at random.
The assumption will be validated by the numerical results of Section V.

\subsection{Keystream Generation}

For the secret matrix $\Sbu$ of Definition~\ref{def:Phinew},
we employ a keystream generator based on a linear feedback shift register (LFSR),
to generate the elements in a fast and efficient manner.
As an example, we introduce
the \emph{self-shrinking generator (SSG)}~\cite{Meier:ssg}.

\begin{df}\label{def:ssg}
Assume that an $L$-stage LFSR
generates a binary $m$-sequence of $\abu = (a_0, a_1, \cdots )$.
With 
a clock-controlled operation,
the self-shrinking generator outputs $b_t = a_{2i+1}$ if $a_{2i} = 1$, and
discards $a_{2i+1}$ if $a_{2i} = 0$.
Then, we obtain a bipolar keystream of
$\sbu = (s_0, s_1, \cdots )$, where $s_t = (-1)^{b_{t}}$ for $t = 0,1, \cdots $,
which will be arranged as the elements of $\Sbu$.
\end{df}

The SSG keystream generation requires a simple structure
of an $L$-stage LFSR along with a clock-controlled operator.
Moreover, the SSG keystream possesses nice pseudorandomness properties~\cite{GolGong:SD}, such as
balance, large period, and large linear complexity.
Meier and Staffelbach~\cite{Meier:ssg} showed that the SSG keystream is balanced, and has the period
of at least $2^{\lfloor L/2 \rfloor}$ and the linear complexity of at least
$2^{\lfloor L/2 \rfloor -1}$, respectively.
Although the SSG keystream generator is considered in this paper,
any other LFSR-based keystream generator can also be applied for the new CS-based cryptosystem.

%

When 
each element of $\Sbu$ is obtained by
a keystream generator,
the initial seed (or state) of the generator is essentially the \emph{key} of the new CS-based cryptosystem.
Therefore, the key should be kept secret between a sender and a legitimate recipient,
while the structure of the keystream generator can be publicly known.



\subsection{AG-OTS Cryptosystem}


\begin{table*}[!t]
\fontsize{9}{11pt}\selectfont
\caption{Symmetric-key AG-OTS Cryptosystem}
\centering
\begin{tabular}{ll}
\hline
\hline
\emph{Public}: & Unitary matrix $\Ubu$, Structure of a keystream generator \\
\emph{Secret}: & Initial seed $\kbu \in \{0, 1\}^L$ of a keystream generator \\
\hline
\emph{Keystream generation}: &  With the initial seed $\kbu$, a keystream generator creates a bipolar keystream of length $MN$. \\
& A secret matrix $\Sbu \in \{-1, +1\}^{M \times N}$ is constructed by arranging the keystream, and  \\
& updated at each encryption by a new keystream.  \\ 
\emph{CS encryption}: & With a plaintext $\xbu \in \R^N$, a ciphertext is produced by $\ybu = \Phibu \xbu \in \R^M $, where $\Phibu = \frac{1}{\sqrt{MN}} \Sbu \Ubu$. \\
\emph{CS decryption}: & Given a noisy ciphertext $\rbu = \Phibu \xbu + \nbu$, the plaintext $\xbu$ is reconstructed by  \\
                      & a CS recovery algorithm with the knowledge of $\Sbu$. \\
\hline
\hline
\end{tabular}
\label{tb:const}
\end{table*}

From Definition~\ref{def:Phinew},
the new CS-based cryptosystem
encrypts a $K$-sparse plaintext $\xbu$\footnote{
In general, $\xbu$ can be $K$-sparse in an arbitrary orthonormal basis $\Psibu $,
i.e., $\xbu = \Psibu\btheta$
with $||\btheta||_0\leq K$, where $\Psibu \neq \frac{1}{\sqrt{N}} \Ubu^T$.
For simplicity, we assume $\Psibu = \Ibu$ in this paper.}
by producing a ciphertext
$\ybu = \Phibu \xbu = \frac{1}{\sqrt{MN}} \Sbu \Ubu \xbu$, 
where $\Sbu$ is updated at each encryption.
Under the presence of noise, 
a legitimate recipient and an adversary have a noisy ciphertext
$\rbu = \Phibu \xbu + \nbu $,
where $\nbu \sim {\mathcal N}({\bf 0}, \sigma^2 \Ibu)$.
As $\Phibu$ is asymptotically Gaussian 
and $\Sbu$ is updated at each encryption,
the new CS-based cryptosystem will be called
the \emph{asymptotically Gaussian one-time sensing (AG-OTS)} cryptosystem throughout this paper.
Table~\ref{tb:const} summarizes the symmetric-key AG-OTS cryptosystem proposed in this paper.

The reliability and stability of the AG-OTS cryptosystem for a legitimate recipient
is straightforward from the RIP result~\cite{Rud:sparse} of a random Gaussian matrix,
under the fact that $\Phibu$ is Gaussian for a sufficiently large $N$. 

\begin{prop}\label{pr:rel}\cite{Rud:sparse}
For a legitimate recipient, if $N$ is sufficiently large,
the AG-OTS cryptosystem theoretically guarantees
a stable and robust CS decryption with bounded errors of a $K$-sparse plaintext,
as long as $M = \mathcal{O}\left( K\log (N/K) \right)$.
\end{prop}


\section{Security Measure}

This section introduces a security measure of the indistinguishability
of a CS-based cryptosystem.
To examine the indistinguishability,
we also discuss the probability metrics of total variation (TV) and Hellinger distances.

\subsection{Indistinguishability}

Assume that
a cryptosystem produces
a ciphertext by encrypting one of two possible plaintexts of the same length.
Then, the cryptosystem is said to have the \emph{indistinguishability}~\cite{Katz:modern}, 
if no adversary can determine in polynomial time
which of the two plaintexts corresponds to the ciphertext,
with probability significantly better than that of a random guess.
In other words,
if a cryptosystem has the indistinguishability,
an adversary is unable to learn any partial information of the plaintext in polynomial time
from a given ciphertext.

\begin{table*}[!t]
\fontsize{9}{11pt}\selectfont
\caption{Indistinguishability Experiment for a CS-based Cryptosystem}
\centering
\begin{tabular}{ll}
\hline
\hline
\emph{Step} 1: & An adversary creates a pair of plaintexts $\xbu_1$ and $\xbu_2$ of the same length, and \\
               & submits them to a CS-based cryptosystem. \\
\emph{Step} 2: & The CS-based cryptosystem encrypts a plaintext $\xbu_h$ by randomly selecting $h \in \{1, 2\}$, and \\
               & gives a noisy ciphertext $\rbu = \Phibu \xbu_h +\nbu $ back to the adversary. \\
\emph{Step} 3: & Given the ciphertext $\rbu$, the adversary carries out  a polynomial time test  ${\mathcal D}: \rbu \rightarrow h' \in \{1, 2\}$, \\
               & to figure out the corresponding plaintext.  \\
\hline
\emph{Decision}: & The adversary passes the experiment if $h' = h$, or fails otherwise.\\
\hline
\hline
\end{tabular}
\label{tb:indist}
\end{table*}

Table~\ref{tb:indist} describes the \emph{indistinguishability experiment}~\cite{Katz:modern}
in the presence of an eavesdropper,
which will be used to investigate the indistinguishability of a CS-based cryptosystem
in this paper.


\subsection{Total Variation (TV) and Hellinger Distances}

In this paper,
we make use of the total variation (TV) distance~\cite{Gibbs:bounding} to evaluate the performance
of an adversary in the indistinguishability experiment of Table~\ref{tb:indist}.
In the experiment, let $d_{\rm TV} (p_1, p_2) $ be the TV distance
between the probability distributions
$p_1 = {\rm Pr}(\rbu | \xbu_1)$ and $p_2 = {\rm Pr}(\rbu | \xbu_2)$.
Then,
it is readily checked from \cite{LeCam:asymp} that
the probability that an adversary can successfully distinguish the plaintexts 
by any kind of test ${\mathcal D}$ is bounded by
\begin{equation}\label{eq:pd}
p_d  
\leq \frac{1}{2} + \frac{d_{\rm TV} (p_1, p_2) }{2}  
\end{equation}
where $d_{\rm TV} (p_1, p_2) \in [0, 1] $.
Therefore, if $d_{\rm TV} (p_1, p_2)$ is zero,
the probability of success
is at most that of a random guess,
which leads to the indistinguishability~\cite{Katz:modern}.

Since computing $d_{\rm TV} (p_1, p_2)$ directly is difficult~\cite{DasGupta:asymp},
we may employ an alternative distance metric to bound the TV distance.
In particular, the \emph{Hellinger} distance~\cite{Gibbs:bounding},
denoted by $d_{\rm H}(p_1, p_2)$, is useful by giving both upper and lower bounds
on the TV distance~\cite{Guntu:sharp}, i.e.,
\begin{equation}\label{eq:bnd_hell}
d_{\rm H} ^2 (p_1, p_2) \leq d_{\rm TV} (p_1, p_2) \leq d_{\rm H}(p_1, p_2) \sqrt{2 - d_{\rm H} ^2 (p_1, p_2)}
\end{equation}
where $d_{\rm H}(p_1, p_2) \in [0, 1]$.
Moreover,
if a ciphertext $\rbu$ conditioned on $\xbu_h$,
is a jointly Gaussian random vector with zero mean and the covariance matrix $\Cbu_h$,
where $h=1$ and $2$,
the Hellinger distance between the multivariate Gaussian distributions $p_1$ and $p_2$
is given by~\cite{Kailath:div}\cite{Karim:note}
\begin{equation}\label{eq:Hell}
d_{\rm H} (p_1, p_2) = \sqrt{ 1- \frac{|\Cbu_1|^{\frac{1}{4}} |\Cbu_2|^{\frac{1}{4}}}{|\Cbu_3|^{\frac{1}{2}}} }
\end{equation}
where $\Cbu_3 = \frac{\Cbu_1 + \Cbu_2}{2}$.
For the formal definitions and properties of the TV and the Hellinger distances,
readers are referred to \cite{Gibbs:bounding}, \cite{LeCam:asymp}, and \cite{DasGupta:asymp}.

Throughout this paper, we use \eqref{eq:pd} $-$ \eqref{eq:Hell}
to examine the success probability of the indistinguishability experiment
for the AG-OTS cryptosystem,
by taking the Gaussian distributed ciphertexts into account.


\section{Security Analysis}
In this section, we show that the AG-OTS cryptosystem can be indistinguishable,
as long as each plaintext has constant energy.
Moreover, we study how much the security of the AG-OTS cryptosystem is sensitive
to energy variation of plaintexts.

\subsection{Indistinguishability}

%

Recall the indistinguishability experiment of Table~\ref{tb:indist}.
Given a plaintext $\xbu_h$,
$ \E [\rbu \arrowvert \xbu_h] = \E[\Phibu ] \xbu_h + \E[\nbu] = \frac{1}{\sqrt{MN}} \E[\Sbu] \cdot \Ubu \xbu_h = {\bf 0}$
from $\E[\Sbu] = {\bf 0}$,
where $h = 1$ and $2$.
In the following,
Lemma~\ref{lm:covN} derives the covariance matrix of $\rbu $
conditioned on $\xbu_h$, by exploiting the independency and the uniformity of the entries of $\Sbu$.

\begin{lem}\label{lm:covN}
In the AG-OTS cryptosystem,
the covariance matrix of $\rbu$ conditioned on $\xbu_h$ is given by
\begin{equation}\label{eq:covN12}
\Cbu_h = \E[\rbu \rbu^T | \xbu_h]
= \left(\frac{|| \xbu_h ||^2}{M} + \sigma ^2 \right) \Ibu
\end{equation}
where $ h = 1$ and $ 2$.
From \eqref{eq:covN12}, it is obvious that
\begin{equation*}\label{eq:covN3}
\Cbu_3 = \frac{\Cbu_1 + \Cbu_2} {2} = \left(\frac{|| \xbu_1||^2 + ||\xbu_2 ||^2}{2M}  + \sigma ^2 \right)\Ibu .
\end{equation*}
\end{lem}

\iproof:
Let $\widehat{\xbu}_h = \Ubu \xbu_h = (\widehat{\xbu}_{h,1}, \cdots, \widehat{\xbu}_{h, N})^T$
for $h = 1$ and $2$, respectively,
where $||\widehat{\xbu}_h||^2 = N ||\xbu_h||^2$.
Also, let $\sbu_k = \Sbu(:, k)$ and $\sbu_l = \Sbu(:, l)$ are the $k$th and
the $l$th column vectors of $\Sbu$, respectively.
Since the elements of $\Sbu$ and $\nbu$ are independent to each other, 
\begin{equation}\label{eq:covN22}
\begin{split}
\Cbu_h  
& = \E \left[ \frac{1}{MN} \Sbu \Ubu \xbu_h \cdot \xbu_h ^T \Ubu^T \Sbu^T \big \arrowvert \xbu_h\right] + \E[\nbu \nbu^T] \\
& = \E \left[ \frac{1}{M N} \sum_{k=1 } ^N \widehat{x}_{h, k} \sbu_k \cdot \sum_{l=1 } ^N \widehat{x}_{h, l}  \sbu_l ^T
\big \arrowvert  \xbu_h \right]  + \sigma ^2 \Ibu \\
& = \frac{1}{MN}  \sum_{k =1 } ^N \sum_{l=1 } ^N  \widehat{x}_{h, k} \widehat{x}_{h, l} \E \left[ \sbu_k \sbu_l ^T \right]
+ \sigma ^2 \Ibu
\end{split}
\end{equation}
where 
\[
\E \left[ \sbu_k \sbu_l ^T \right]
= \left\{ \begin{array}{ll} \Ibu, & \quad \mbox{ if } k = l, \\
{\bf 0}, & \quad \mbox{ if } k \neq l \end{array} \right.
\]
as the entries of $\sbu_k $ and $\sbu_l $ take $\pm 1$ independently and uniformly at random.
Thus, \eqref{eq:covN22} yields
\begin{equation*}\label{eq:covN122}
\begin{split}
\Cbu_h & =  \frac{1}{MN} \left(\sum_{k =1 } ^N  \widehat{x}_{h, k} ^2 \right) \cdot \Ibu + \sigma ^2 \Ibu \\
& = \left(\frac{|| \xbu_h ||^2}{M}  +  \sigma ^2 \right)\Ibu
\end{split}
\end{equation*}
which completes the proof.
\qed

In Lemma~\ref{lm:covN}, note that
the derivation of covariance matrices
does not rely on the asymptotic Gaussianity of $\Phibu$.
Instead, the covariance matrices are non-asymptotic results, obtained by
exploiting the independency and the uniformity of the elements of $\Sbu$.

Using the covariance matrices of Lemma~\ref{lm:covN},
we can develop upper and lower bounds on the TV distance in the AG-OTS cryptosystem,
which is the main contribution of this paper.

\begin{thr}\label{th:tv_bnd2}
In the AG-OTS cryptosystem,
assume that the plaintext length $N$ is sufficiently large
such that $\Phibu$ can be asymptotically Gaussian by Theorem~\ref{th:aGauss}.
In the indistinguishability experiment, 
let $d_{\rm TV} (p_1, p_2)$ be the TV distance between probability distributions
of ciphertexts conditioned on a pair of plaintexts in the AG-OTS cryptosystem.
Let $\xbu_{\min}$ and $\xbu_{\max}$ be the plaintexts
that have the minimum and maximum possible energies, respectively,
where $\gamma = \frac{||\xbu_{\min} ||^2}{||\xbu_{\max} ||^2}$ is
the minimum energy ratio and
${\rm PNR}_{\max} = \frac{||\xbu_{\max}||^2}{M \sigma ^2}$
is the maximum plaintext-to-noise power ratio, respectively,
of the cryptosystem.
Then, the worst-case lower and upper bounds on $d_{\rm TV} (p_1, p_2)$
are given by
\begin{equation}\label{eq:tv_bnd2}
\begin{split}
& d_{\rm TV, low} = 1 - \left( \frac{4 \gamma_e}{(\gamma_e+1)^2} \right)^{\frac{M}{4}}, \\
& d_{\rm TV, up} = \sqrt{ 1 - \left( \frac{4 \gamma_e}{(\gamma_e+1)^2} \right)^{\frac{M}{2}}},
\end{split}
\end{equation}
respectively, where
\begin{equation}\label{eq:gam_e}
\gamma_e 
= \frac{1+ \gamma \cdot {\rm PNR}_{\max}}{ 1+ {\rm PNR}_{\max}}.
\end{equation}
\end{thr}

\iproof:
In the indistinguishability experiment of Table~\ref{tb:indist}, let us consider a pair of
plaintexts $\xbu_1$ and $\xbu_2 $,
where ${\rm PNR}_h = \frac{||\xbu_h||^2}{M \sigma ^2}$ for $h=1$ and $2$.
From the covariance matrices of Lemma~\ref{lm:covN},
\begin{equation*}\label{eq:detCA}
|\Cbu_h| =  \left(\frac{||\xbu_h ||^2}{M} + \sigma ^2 \right)^M
 = \sigma ^{2M} \cdot \left( {\rm PNR}_h  + 1 \right)^M
\end{equation*}
for each $h$. Obviously,
\begin{equation*}\label{eq:detCA3}
|\Cbu_3| =  \sigma ^{2M} \cdot  \left( \frac{{\rm PNR}_1 + {\rm PNR}_2}{2}  + 1 \right)^M.
\end{equation*}
In~\eqref{eq:Hell},
\begin{equation*}\label{eq:gammaW}
\begin{split}
\Gamma  = \frac{| \Cbu_1 |^{\frac{1}{4}} \cdot | \Cbu_2 |^{\frac{1}{4}}}{| \Cbu_3 |^{\frac{1}{2}}}
& = \left( \frac{( {\rm PNR}_1  + 1)( {\rm PNR}_2  + 1)}
{ \left(\left(\frac{{\rm PNR}_1 + {\rm PNR}_2}{2} \right)  + 1 \right)^2} \right)^{\frac{M}{4}} \\
& = \left(  \frac{4 \gamma_e }{(\gamma_e+1)^2} \right)^{\frac{M}{4}}
\end{split}
\end{equation*}
where
\[
\gamma_e = \frac{1+  {\rm PNR}_1}{ 1+ {\rm PNR}_2}
= \frac{1+ \gamma \cdot {\rm PNR}_2}{ 1+ {\rm PNR}_2}
\]
and $\gamma = \frac{||\xbu_1 ||^2}{||\xbu_2 ||^2}$.
With $d_{\rm H} (p_1, p_2) = \sqrt{1 - \Gamma}$,
\eqref{eq:bnd_hell} yields the lower and upper bounds of the form of \eqref{eq:tv_bnd2}.
Without loss of generality,
we may assume $|| \xbu_1 ||^2 \leq || \xbu_2||^2$, which yields $0 \leq \gamma \leq 1$.
As the lower and upper bounds turn out to be monotonically decreasing over $\gamma \in [0, 1]$,
we can redefine $\gamma = \frac{||\xbu_{\min}||^2}{||\xbu_{\max}||^2}$
and $\gamma_e = \frac{1+ \gamma \cdot {\rm PNR}_{\max}}{ 1+ {\rm PNR}_{\max}}$, $0 \leq  \gamma_e \le 1$
with $\xbu_1 = \xbu_{\min} $ and $\xbu_2 = \xbu_{\max}$,
to obtain the worst-case bounds,
which completes the proof.
\qed

In~\eqref{eq:gam_e},
$\gamma_e$ is a general definition of the energy ratio
covering noisy cases,
which will be called the \emph{effective} energy ratio in this paper.
For security analysis, we assume that
both a legitimate recipient and an adversary have
the same energy ratio $\gamma$ and the same ${\rm PNR}_{\max}$
in the AG-OTS cryptosystem.

Theorem~\ref{th:tv_bnd2} shows that
the indistinguishability of the AG-OTS cryptosystem depends on the ciphertext length $M$,
the minimum energy ratio $\gamma$, and the maximum plaintext-to-noise ratio ${\rm PNR}_{\max}$,
irrespective of the plaintext length $N$ and
the sparsity $K$.
In particular, if $\gamma = \gamma_e = 1$, 
the indistinguishability can be guaranteed for the AG-OTS cryptosystem,
regardless of $M$ and ${\rm PNR}_{\max}$.


\begin{cor}\label{co:perfA}
If each plaintext has constant energy 
or $\gamma = 1$,
the AG-OTS cryptosystem has the indistinguishability,
since the success probability of the indistinguishability experiment is at most $0.5$
from~\eqref{eq:pd},
thanks to $d_{\rm TV} (p_1, p_2) = 0$ for $ d_{\rm TV, low} = d_{\rm TV, up} = 0$.
\end{cor}

In the AG-OTS cryptosystem,
Corollary~\ref{co:perfA} ensures that
no adversary can learn any partial information about the plaintext from a given ciphertext, 
as long as each plaintext has constant energy,
which is also the case in the G-OTS cryptosystem of~\cite{Bianch:linear}.
To achieve the indistinguishability, therefore,
a normalization step for equalizing the plaintext energy 
is implicitly required before CS encryption in the AG-OTS cryptosystem of Table~\ref{tb:const}.
Since it also offers a practical benefit from the efficient keystream generation,
the AG-OTS cryptosystem can be a promising option for information security, 
by guaranteeing the indistinguishability, reliability, and efficiency in a CS framework.

\subsection{Energy Sensitivity}

\begin{figure}[!t]
\centering
\includegraphics[width=0.48\textwidth, angle=0]{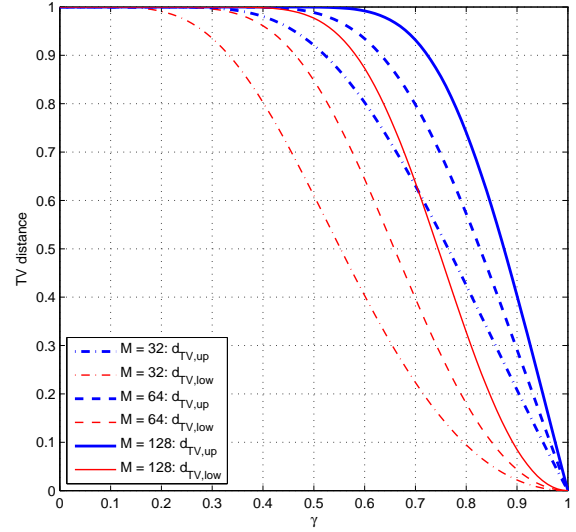}
\caption{Upper and lower bounds on the TV distance over $\gamma$ in the noiseless AG-OTS cryptosystem.}
\label{fig:dtv_uplow}
\end{figure}

Theorem~\ref{th:tv_bnd2} implies that
the indistinguishability of the AG-OTS cryptosystem
can be \emph{sensitive} to the minimum energy ratio $\gamma$. 
Figure~\ref{fig:dtv_uplow} sketches the upper and lower bounds of \eqref{eq:tv_bnd2}
over $ \gamma$ at ${\rm PNR}_{\max} = \infty$ in the noiseless AG-OTS cryptosystem.
It indicates that the TV distance increases as
$\gamma$ gets away from $1$.
In particular, if $M$ gets larger,
the TV distance approaches to $1$ more quickly as $\gamma$ decreases.
Such a behavior of the TV distance 
suggests that if $\gamma$ is far less than $1$,
an adversary may be able to detect a correct plaintext in the indistinguishability experiment
with a significantly high probability of success,
which implies that
the AG-OTS cryptosystem may not be indistinguishable. 


\begin{figure}[!t]
\centering
\includegraphics[width=0.48\textwidth, angle=0]{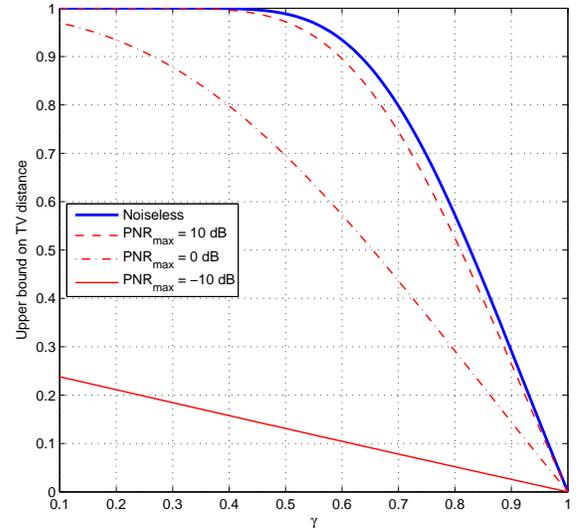}
\caption{Upper bounds on the TV distance over $\gamma$ in the noisy AG-OTS cryptosystem,
where $M = 64$.}
\label{fig:dtv_uplow_e}
\end{figure}

In addition, Figure~\ref{fig:dtv_uplow_e} shows the upper bounds of \eqref{eq:tv_bnd2}
over $\gamma$ for various ${\rm PNR_{\max}}$ in the noisy AG-OTS cryptosystem,
where $M=64$.
In the figure,
the bounds are sensitive to $\gamma$ for each ${\rm PNR}_{\max}$,
as in the noiseless case of Figure~\ref{fig:dtv_uplow}.
Moreover,
the bound itself is smaller at less ${\rm PNR}_{\max}$,
which implies that an adversary may have a difficulty
in distinguishing plaintexts at low ${\rm PNR}_{\max}$, 
due to the low TV distance.
As a result, it appears that the security of the AG-OTS cryptosystem would be
more sensitive to the energy ratio $\gamma$ at higher ${\rm PNR}_{\max}$.

%


In summary,
the AG-OTS cryptosystem
may not be able to achieve the indistinguishability,
unless each plaintext has constant energy.
In what follows,
we study how much the security of the AG-OTS cryptosystem is sensitive
to energy variation of plaintexts.
It is worth studying
the energy sensitivity, 
since
one might need to assign unequal energy for each plaintext in the presence of noise,
depending on the reliability demands.
Theorems~\ref{th:sensM} $-$ \ref{th:pnr_max} present
sufficient conditions for the minimum energy ratio $\gamma$, the plaintext length $M$,
and the maximum plaintext-to-noise ratio ${\rm PNR}_{\max}$, respectively,
to guarantee the \emph{asymptotic} indistinguishability for the AG-OTS cryptosystem. 

%

\begin{thr}\label{th:sensM}
When $M$ and ${\rm PNR}_{\max}$ are given,
let $\varphi = (1- 4 \epsilon_N ^2)^{-\frac{2}{M}}$
and $\gamma_{e, \min} =2 \varphi - 1 - 2 \sqrt{\varphi(\varphi-1)}$, respectively.
If the minimum energy ratio $\gamma$ satisfies
\begin{equation}\label{eq:gam}
\gamma \ge \gamma_{e, \min} - (1 - \gamma_{e, \min}) \cdot {\rm PNR}_{\max} ^{-1}
\triangleq \gamma_{\min},
\end{equation}
the success probability of the indistinguishability experiment is
$p_d \leq 0.5+\epsilon_N $,
where $\epsilon_N$ vanishes as the plaintext length $N$ increases.
In other words, the AG-OTS cryptosystem is asymptotically indistinguishable
for a sufficiently large $N$,
as long as $\gamma \ge \gamma_{\min}$ 
for given $M$ and ${\rm PNR}_{\max}$.
\end{thr}

\iproof:
When $M$ is given,
we have
$ \frac{4\gamma_e}{(\gamma_e+1)^2} \ge \left(1 - 4 \epsilon_N ^2 \right)^{\frac{2}{M}}$
from $d_{\rm TV, up} \leq 2\epsilon_N$.
The inequality turns into
$ \gamma_e ^2 - 2(2 \varphi -1) \gamma_e + 1 \leq 0$,
which holds if $\gamma_e \geq \gamma_{e, \min}$,
or equivalently if $\gamma \ge \gamma_{\min} $ in \eqref{eq:gam}.
Consequently, if the sufficient condition of \eqref{eq:gam} is met, 
the success probability of the indistinguishability experiment is
$p_d \leq 0.5 + \epsilon_N$ by \eqref{eq:pd},
which completes the proof.
\qed

%

\begin{figure}[!t]
\centering
\includegraphics[width=0.48\textwidth, angle=0]{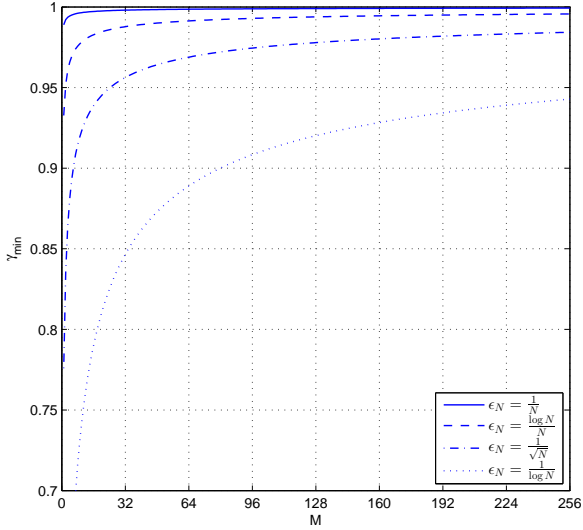}
\caption{Minimum energy ratio required for the asymptotic indistinguishability of
the AG-OTS cryptosystem at ${\rm PNR}_{\max}=20$ dB, where $N = 512$.}
\label{fig:gam_min}
\end{figure}

In Theorem~\ref{th:sensM},
$\gamma_{\min}$ is the minimum energy ratio required
for the asymptotic indistinguishability of the AG-OTS cryptosystem.
Figure~\ref{fig:gam_min} displays $\gamma_{\min}$ over $M$
in the AG-OTS cryptosystem at ${\rm PNR}_{\max} = 20$ dB,
where $N = 512$. 
In the figure, $\gamma_{\min}$ is sketched
for various $\epsilon_N = \frac{1}{\log N}, \frac{1}{\sqrt{N}}, \frac{\log N}{N}, $
and $\frac{1}{N}$.
The figure reveals that the minimum energy ratio required
for the asymptotic indistinguishability
approaches to $1$ as the ciphertext length $M$ increases.
In particular, if the AG-OTS cryptosystem allows larger energy variation for plaintexts,
the asymptotic indistinguishability can be achieved at a lower rate over $N$.

\begin{thr}\label{th:sensG}
When $\gamma$ and ${\rm PNR}_{\max}$ are given,
recall $\gamma_e = \frac{1+ \gamma \cdot {\rm PNR}_{\max}}{ 1+ {\rm PNR}_{\max}}$.
Let $C_{\gamma_e} = \log \frac{4 \gamma_e}{(\gamma_e+1)^2} \leq 0$,
where the equality holds if and only if $\gamma=1$.
If the ciphertext length $M$ satisfies
\begin{equation*}\label{eq:fM}
M \leq \frac{2}{C_{\gamma_e}} \log (1 - 4 \epsilon_N ^2)
\triangleq M_{\max} ,
\end{equation*}
then $p_d \leq 0.5 + \epsilon_N$, which implies that
the AG-OTS cryptosystem is asymptotically indistinguishable for a sufficiently large $N$,
as long as $M \leq M_{\max} $ 
for given $\gamma$ and ${\rm PNR}_{\max}$.
\end{thr}
\iproof:
When $\gamma_e$ is given from $\gamma$ and ${\rm PNR}_{\max}$,
the proof is similar to that of Theorem~\ref{th:sensM}
from $d_{\rm TV, up} \leq 2 \epsilon_N$.
\qed

\begin{figure}[!t]
\centering
\includegraphics[width=0.48\textwidth, angle=0]{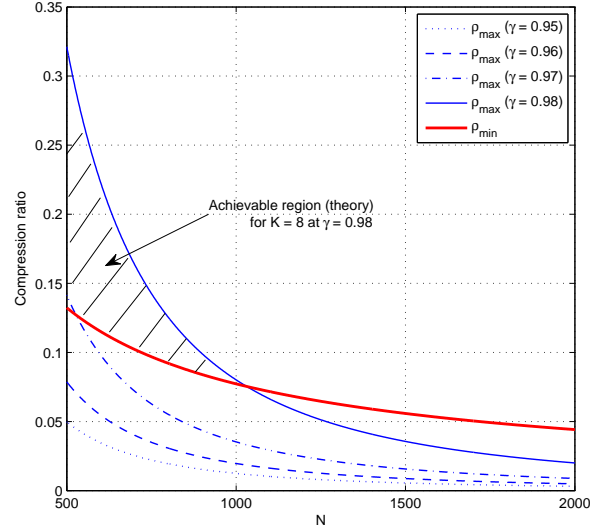}
\caption{Compression ratios for the asymptotic indistinguishability and the reliability of the AG-OTS cryptosystem at ${\rm PNR}_{\max}=20$ dB,
where $K=8$ and $\epsilon_N = 1/{\sqrt{N}}$.}
\label{fig:rho_max}
\end{figure}

Figure~\ref{fig:rho_max} depicts the maximum compression ratio $\rho_{\max} = \frac{M_{\max}}{N}$ over $N$
for the AG-OTS cryptosystem to be asymptotically indistinguishable
at ${\rm PNR}_{\max} = 20 $ dB
with $p_d \leq 0.5+ \epsilon_N$,
where $\epsilon_N = \frac{1}{\sqrt{N}}$. 
We also sketch the minimum compression ratio $\rho_{ \min} = \frac{2 K \log(N/K)}{N}$ for
reliable CS decryption\footnote{This is a theoretical ratio in noiseless recovery.}
from a random Gaussian sensing~\cite{Foucart:mathCS} with $K=8$,
to compare the requirements for the asymptotic indistinguishability and the reliability.
Note that if $\gamma= 1$ or the plaintexts have constant energy,
the indistinguishability can be achieved at any compression ratio.
Meanwhile, if $\gamma < 1$, 
the compression ratio of the AG-OTS cryptosystem
must be at most $\rho_{\max}$ 
for the asymptotic indistinguishability.
In particular, if $\rho_{ \max} < \rho_{ \min}$,
the cryptosystem may not be valid at least in theory for the corresponding $N$, since
the indistinguishability cannot be compatible with the reliability.
Thus, Figure~\ref{fig:rho_max} shows that if $\gamma = 0.98$,
the AG-OTS cryptosystem at ${\rm PNR}_{\max} = 20 $ dB can achieve both reliability and security
for the plaintexts of at most $K = 8$ nonzero entries
only at the compression ratios of the achievable (shaded) region.
It also shows that if $\gamma \leq 0.96$,
the AG-OTS cryptosystem has no theoretically achievable region 
for $N>500$,
where the reliability and the indistinguishability cannot be guaranteed simultaneously.

%


In \eqref{eq:gam_e}, note that
\begin{equation}\label{eq:gmin_ge}
\gamma_e = \gamma + \frac{1-\gamma}{1+ {\rm PNR}_{\max}}
\end{equation}
where
$\gamma_e \ge  \gamma$ for $\gamma \in [0,1]$.
Since 
the upper bound of \eqref{eq:tv_bnd2}
is monotonically decreasing over $\gamma_e \in [0, 1]$,
\eqref{eq:gmin_ge} implies that
the upper bound on the TV distance 
is lower in noisy case (${\rm PNR}_{\max} < \infty$)
than in noiseless case (${\rm PNR}_{\max} = \infty$).
Ultimately, it points out that
the presence of noise improves the security of the AG-OTS cryptosystem
by lowering the success probability of an adversary in the indistinguishability experiment.
Moreover,  
one can increase $\gamma_e$
by reducing ${\rm PNR}_{\max}$ in \eqref{eq:gmin_ge} for a given $\gamma$,
which indicates that
the AG-OTS cryptosystem will be more secure for less ${\rm PNR}_{\max}$.
With given $\gamma$ and $M$,
Theorem~\ref{th:pnr_max}
presents the largest possible ${\rm PNR}_{\max}$
to guarantee the asymptotic indistinguishability
for the AG-OTS cryptosystem,
where the proof is straightforward from
$\gamma_e \ge \gamma_{e, \min}$ in \eqref{eq:gmin_ge}.

\begin{thr}\label{th:pnr_max}
In the AG-OTS cryptosystem,
assume that
the minimum energy ratio 
is given as $\gamma < \gamma_{e, \min}$ for a given $M$,
where $\gamma_{e, \min}$ is the minimum effective energy ratio 
defined in Theorem~\ref{th:sensM}.
Then, the asymptotic indistinguishability can be achieved for a sufficiently large $N$,
if 
\begin{equation*}\label{eq:pnr_max}
{\rm PNR}_{\max} \leq \frac{1 - \gamma_{e, \min}}{\gamma_{e, \min} - \gamma}.
\end{equation*}
\end{thr}


\begin{figure}[!t]
\centering
\includegraphics[width=0.48\textwidth, angle=0]{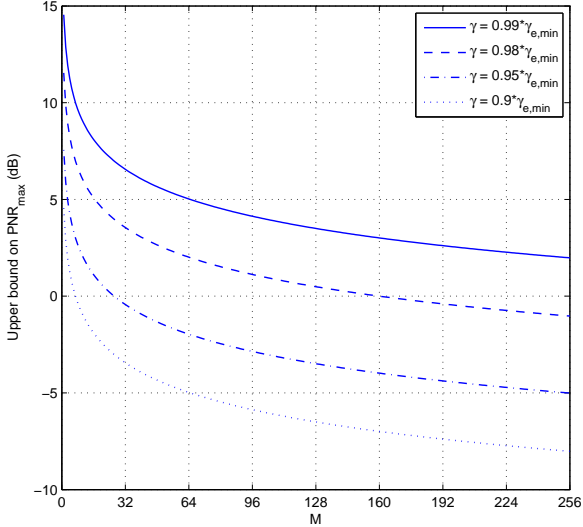}
\caption{Upper bounds on ${\rm PNR}_{\max}$ for the asymptotic indistinguishability of the AG-OTS cryptosystem,
where $N = 512$ and $\epsilon_N = 1/{\sqrt{N}}$.}
\label{fig:pnr_max}
\end{figure}

Note that if $\gamma \ge \gamma_{e, \min}$, the AG-OTS cryptosystem
is asymptotically indistinguishable, regardless of ${\rm PNR}_{\max}$,
due to $\gamma_e \ge \gamma \ge \gamma_{e, \min}$.
Figure~\ref{fig:pnr_max} displays the upper bounds
on ${\rm PNR}_{\max}$ of Theorem~\ref{th:pnr_max} for various $\gamma < \gamma_{e, \min}$, 
where $N = 512$ and $\epsilon_N = \frac{1}{\sqrt{N}}$.
From \eqref{eq:gmin_ge},
it is clear that if ${\rm PNR}_{\max}$ is sufficiently high,
$\gamma_e \approx \gamma < \gamma_{e,\min}$
from which the asymptotic indistinguishability cannot be achieved from Theorem~\ref{th:sensM}.
Figure~\ref{fig:pnr_max} points out that
we need to increase $\gamma_e$ by reducing ${\rm PNR}_{\max}$ below the upper bound for each given $\gamma$,
to achieve the asymptotic indistinguishability of the AG-OTS cryptosystem.
However, it appears that the largest possible ${\rm PNR}_{\max}$ is relatively low
for a reliable CS decryption.
For the AG-OTS cryptosystem, therefore,
it is an important issue to keep the energy variation of plaintexts as low as possible.

In conclusion, it turned out that the security of the AG-OTS cryptosystem is highly sensitive to the energy ratio
of plaintexts.
The indistinguishability can be achieved only if 
all the plaintexts have equal and constant energy.
Therefore,
if the AG-OTS cryptosystem is to be indistinguishable non-asymptotically,
it is essential that each plaintext should be normalized before CS encryption to have constant energy.
By analyzing the energy sensitivity,
we presented the sufficient conditions of Theorems~\ref{th:sensM} $-$ \ref{th:pnr_max}
for the asymptotic indistinguishability of the AG-OTS cryptosystem with unequal plaintext energy.
However, we found that even the asymptotic indistinguishability
can be achieved 
only if the plaintexts have low energy variation for most $M$, $N$, and ${\rm PNR}_{\max}$. 
As the analysis technique utilizes the result of Theorem~\ref{th:tv_bnd2}
based on the Gaussianity of the sensing matrix,
the energy sensitivity of this paper can also be valid for the G-OTS cryptosystem,
which has never been discussed in~\cite{Bianch:linear}.

\section{Numerical Results}

This section presents numerical results to demonstrate the indistinguishability
and the energy sensitivity of the AG-OTS cryptosystem.
In numerical experiments,
each plaintext $\xbu$ has at most $K  $ nonzero entries,
where the positions are chosen uniformly at random and
the coefficients are taken from the Gaussian distribution.
In CS encryption,
$\Phibu= \frac{1}{\sqrt{MN}} \Sbu \Ubu$,
where $\Ubu = \Dbu$ is the discrete cosine transform (DCT) matrix.
Each element of the secret matrix $\Sbu$ is taken from 
a bipolar keystream
obtained by
the self-shrinking generator (SSG) with a $128$-stage LFSR.
For comparison, we test with $\Sbu$ whose elements are taken from
the random Bernoulli distribution.
We assume that a ciphertext is available for
both an adversary and a legitimate recipient
with the same 
${\rm PNR} = \frac{|| \xbu ||^2}{M \sigma ^2}$.
For CS decryption,
the CoSaMP recovery algorithm~\cite{Needell:cosamp} is employed for a legitimate recipient
to decrypt each ciphertext
with the knowledge of $\Sbu$.
Meanwhile, we assume that
an adversary can attempt any kind of detection in polynomial time,
to pass the indistinguishability experiment
by distinguishing a pair of plaintexts from a given ciphertext.



\begin{figure*}[!t]
\centering
\includegraphics[width=0.88\textwidth, angle=0]{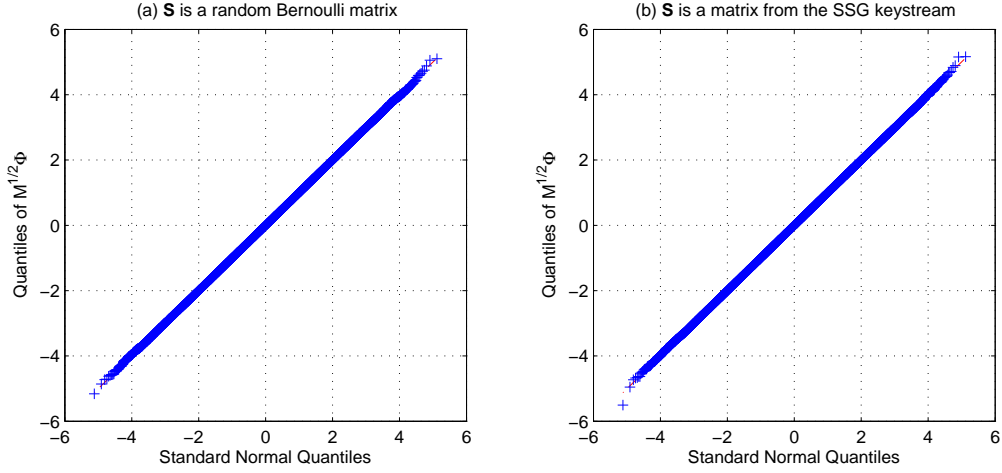}
\caption{Quantile-Quantile plots (QQ-plots) of the entries of total $100$ matrices of $\sqrt{M} \Phibu$'s in the AG-OTS cryptosystem,
where $N=512$ and $M=64$.}
\label{fig:qqplot}
\end{figure*}

\begin{figure*}[!t]
\centering
\includegraphics[width=0.88\textwidth, angle=0]{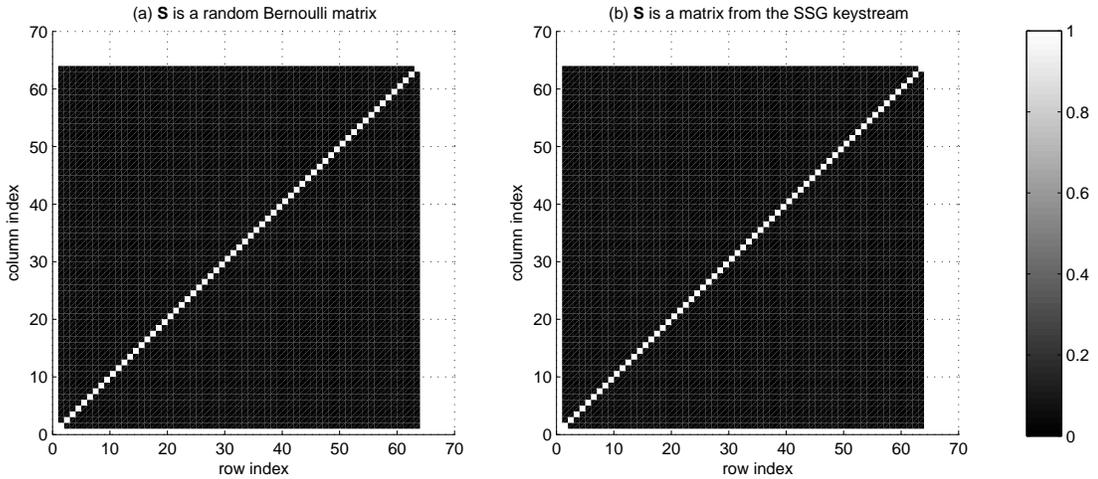}
\caption{Covariance matrices of $M \cdot \E[\rbu \rbu^T | \xbu]$ in the AG-OTS cryptosystem at ${\rm PNR} = 20 $ dB,
where $N=512$, $M=64$, and $K=8$. For a given $\xbu$ with $|| \xbu||^2=1$, total $10000$ matrices of $\Sbu$ have been tested for the average.}
\label{fig:cov}
\end{figure*}

Figure~\ref{fig:qqplot} displays the quantile-quantile (QQ) plots
of the entries of total $100$ matrices of $\sqrt{M} \Phibu$
in the AG-OTS cryptosystem, where $N = 512$ and $M=64$.
In Figure~\ref{fig:qqplot}(a), each entry of $\Sbu$ is taken from the random Bernoulli distribution
taking $\pm 1$ independently and uniformly at random, while
$\Sbu$ of Figure~\ref{fig:qqplot}(b) is from the bipolar SSG keystream.
Since both QQ-plots are linear with slope $1$, it appears that
the entries of $\sqrt{M} \Phibu$ follow the normal distribution 
in both cases of $\Sbu$.
The figure gives a numerical evidence that $\Phibu$ of the
AG-OTS cryptosystem is asymptotically Gaussian for a sufficiently large $N$,
even if $\Sbu$ is generated in a pseudorandom fashion by the SSG.

Figure~\ref{fig:cov} illustrates the covariance matrices of
$M \Cbu = M \cdot \E[\rbu \rbu ^T | \xbu]$ in the AG-OTS cryptosystem at ${\rm PNR} = 20 $ dB,
where 
$N=512$, $M=64$, and $K=8$.
In the experiment,
total $10000$ matrices of $\Sbu$ have been tested for the average
with a given $\xbu$ of $|| \xbu||^2=1$.
In the figure,
the dark areas indicate the off-diagonal entries of each covariance matrix having very small magnitudes
less than $0.04$,
whereas the white cells represent the diagonal components of significant values,
determined by the plaintext energy and the noise variance.
Figure~\ref{fig:cov} numerically confirms that
the covariance analysis of Lemma~\ref{lm:covN} is valid for the AG-OTS cryptosystem,
whether $\Sbu$ is a random Bernoulli matrix or a matrix from the SSG keystream.

\begin{figure}[!t]
\centering
\includegraphics[width=0.48\textwidth, angle=0]{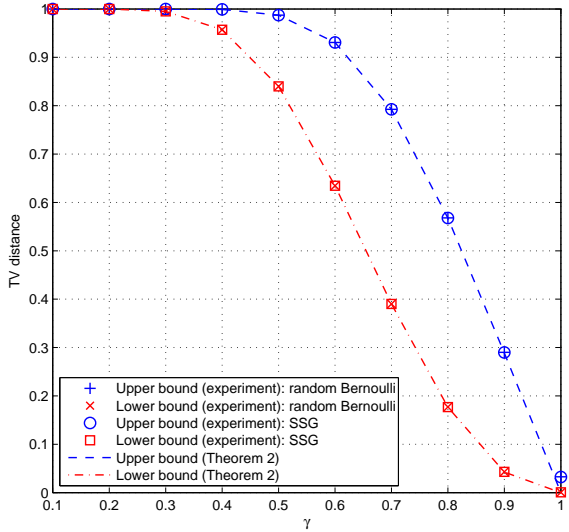}
\caption{Upper and lower bounds on TV distance over $\gamma$ in the AG-OTS cryptosystem at ${\rm PNR}_{\max} = 20$ dB,
where $N=512$, $M=64$, and $K=8$. For a given $\xbu$, total $10^6$ matrices of $\Sbu$ have been tested.}
\label{fig:dtv_AG}
\end{figure}

Figure~\ref{fig:dtv_AG} displays the upper and lower bounds of Theorem~\ref{th:tv_bnd2}
on the TV distance over $\gamma$ in the AG-OTS cryptosystem at ${\rm PNR}_{\max} = 20$ dB,
where $N=512$, $M=64$, and $K=8$.
In the experiment, we computed
the bounds of \eqref{eq:bnd_hell} using the covariance matrices obtained by testing
total $10^6$ matrices of $\Sbu$,
where each entry of $\Sbu$ is taken from
the random Bernoulli distribution or the SSG keystream.
In both cases of $\Sbu$,
the figure shows that the bounds from the experiment
are well matched to the theoretical results of Theorem~\ref{th:tv_bnd2}.
In summary, Figures~\ref{fig:qqplot}$-$\ref{fig:dtv_AG} validate
our assumption of the independency and the uniformity of the elements of $\Sbu$ from the SSG keystream
through the numerical experiments.


\begin{figure}[!t]
\centering
\includegraphics[width=0.48\textwidth, angle=0]{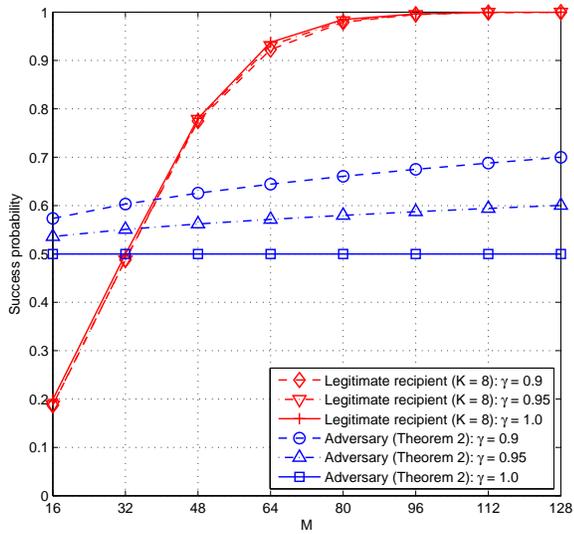}
\caption{Success probabilities over $M$ in the AG-OTS cryptosystem at ${\rm PNR}_{\max} = 20$ dB,
where $N=512$. For an adversary, the upper bounds on the success probability of the indistinguishability experiment are sketched.}
\label{fig:succ}
\end{figure}

Figure~\ref{fig:succ} displays the success probabilities over the ciphertext length $M$
in the AG-OTS cryptosystem at ${\rm PNR}_{\max} = 20$ dB,
where $N=512$.
For an adversary, it sketches the upper bounds on the success probability of the indistinguishability experiment,
obtained by \eqref{eq:pd} from the upper bound of Theorem~\ref{th:tv_bnd2}.
For comparison,
we also sketch the empirical success probabilities 
of a legitimate recipient, where
we tested total $10000$ plaintexts each of which has at most $K=8$ nonzero entries
and the energy $ ||\xbu||^2 = \alpha ||\xbu_{\max} ||^2 $
with $\alpha$ uniformly distributed in $[\gamma, 1]$.
In CS encryption, each entry of the secret matrix $\Sbu$ is from the SSG keystream,
where we observed that the decryption performance
is similar to that of $\Sbu$ from the random Bernoulli distribution.
The CS decryption is declared as a success if a decrypted plaintext $\widehat{\xbu}$ achieves
$\frac{||\xbu - \widehat{\xbu}||^2}{||\xbu||^2} < 10^{-2}$.
The figure shows that
a legitimate recipient enjoys a reliable and stable CS decryption for a sufficiently large $M$
at each $\gamma$.
Meanwhile, the upper bounds on
the success probability of an adversary 
indicate that
no detection test can be successful in
the indistinguishability experiment
with the probability more than the bounds.
In particular, if $\gamma = 1$,
no adversary can learn any information about the plaintext
with the success probability higher than $0.5$,
which leads to the indistinguishability. 
However, if energy variation occurs in plaintexts with $\gamma < 1$,
the figure reveals that
an adversary may be able to distinguish the plaintexts in the experiment, with the probability higher than $0.5$.
It also shows that the success probability of an adversary becomes more significant as the minimum energy ratio $\gamma$ decreases
and the plaintext length $M$ increases.

\begin{figure}[!t]
\centering
\includegraphics[width=0.48\textwidth, angle=0]{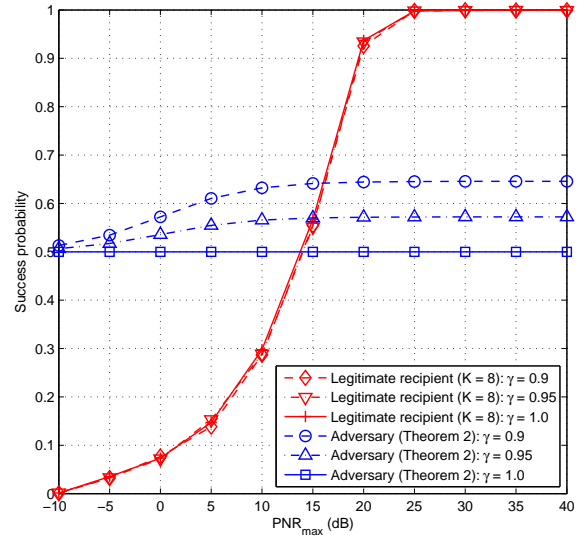}
\caption{Success probabilities over ${\rm PNR}_{\max}$ in the AG-OTS cryptosystem,
where $N=512$ and $M=64$. For an adversary, the upper bounds on the success probability of the indistinguishability experiment are sketched.}
\label{fig:succ2}
\end{figure}

Figure~\ref{fig:succ2} depicts the success probabilities over ${\rm PNR}_{\max}$ in the AG-OTS cryptosystem,
where $N=512$ and $M=64$. 
The simulation environment is identical to that of Figure~\ref{fig:succ}.
As can be seen from the figure,
the CS decryption performance of a legitimate recipient improves over ${\rm PNR}_{\max}$.
However, the detection performance of an adversary
is saturated at high ${\rm PNR}_{\max}$,
where the highest possible success probability 
is determined by the minimum energy ratio $\gamma$.
The figure also shows that if ${\rm PNR}_{\max}$ is low,
the highest possible success probability of an adversary is close to $0.5$
for any $\gamma$,
which implies that the AG-OTS cryptosystem can be indistinguishable at sufficiently low ${\rm PNR}_{\max}$,
regardless of energy variation.
In this case, however, a legitimate recipient also fails in CS decryption due to high noise level.

%
%

\section{Conclusions}

This paper has proposed a new CS-based cryptosystem,
named as the AG-OTS cryptosystem, by employing a secret bipolar keystream
and a public unitary matrix
for efficient implementation in practice.
We demonstrated that the elements of the sensing matrix are asymptotically Gaussian
for a sufficiently large plaintext length,
which guarantees
a stable and robust CS decryption for a legitimate recipient.
By means of the total variation (TV) and the Hellinger distances,
we showed that
the AG-OTS cryptosystem can have the indistinguishability against an adversary,
as long as each plaintext has constant energy.
Therefore, it is essential that the AG-OTS cryptosystem should have a normalization step before CS encryption
for equalizing the plaintext energy, which guarantees the computational security against
any kind of polynomial time attack from an adversary.
Finally, we found that the indistinguishability of the AG-OTS cryptosystem
is highly sensitive to energy variation of plaintexts.
To support the AG-OTS cryptosystem with unequal plaintext energy,
we developed sufficient conditions on the minimum energy ratio, the plaintext length,
and the maximum plaintext-to-noise power ratio,
respectively, for the asymptotic indistinguishability.
The results of the energy sensitivity 
can be directly applicable to the G-OTS cryptosystem of~\cite{Bianch:linear}.
\ifCLASSOPTIONcaptionsoff
\newpage
\fi

\end{document}